# Experimental Study of Heat Pump Thermodynamic Cycles Using $CO_2$ Based Mixtures - Methodology and First Results

Paul Bouteiller [a)], Marie-France Terrier [b)], Pascal Tobaly [c)]

*CNAM Laboratory CMGPCE*
*292 rue Saint Martin – case 2D3P21, Paris, 75003, FRANCE*

[a)] paul.bouteiller@cnam.fr
[b)] marie-france.terrier@lecnam.net
[c)] Corresponding author: pascal.tobaly@lecnam.net

**Abstract.** The aim of this work is to study heat pump cycles, using $CO_2$ based mixtures as working fluids. Since adding other chemicals to $CO_2$ moves the critical point and generally equilibrium lines, it is expected that lower operating pressures as well as higher global efficiencies may be reached. A simple stage pure $CO_2$ cycle is used as reference, with fixed external conditions. Two scenarios are considered: water is heated from 10 °C to 65 °C for Domestic Hot Water scenario and from 30 °C to 35 °C for Central Heating scenario. In both cases, water at the evaporator inlet is set at 7 °C to account for such outdoor temperature conditions. In order to understand the dynamic behaviour of thermodynamic cycles with mixtures, it is essential to measure the fluid circulating composition. To this end, we have developed a non intrusive method. Online optical flow cells allow the recording of infrared spectra by means of a Fourier Transform Infra Red spectrometer. A careful calibration is performed by measuring a statistically significant number of spectra for samples of known composition. Then, a statistical model is constructed to relate spectra to compositions. After calibration, compositions are obtained by recording the spectrum in few seconds, thus allowing for a dynamic analysis. This article will describe the experimental setup and the composition measurement techniques. Then a first account of results with pure $CO_2$, and with the addition of propane or R-1234yf will be given.

## 1. INTRODUCTION

International protocols and regulations are currently implemented to reduce the impact of human activities on Earth's climate. The objective is reduction of the greenhouse gas effect through limitation of the release in the atmosphere of involved gases. The revision of the European F-gas regulation (2014) establishes quotas for many HFC refrigerants, depending on their Global Warming Potential (GWP). New HFO refrigerants are designed to reduce GWP of synthetic refrigerants, and natural refrigerants are becoming more attractive for commercial refrigeration and heat pumps.

In order to improve the efficiency of heat pumps in central heating production mode, we currently experiment the use of $CO_2$ based mixtures as refrigerants, with measurements of the circulating fluid composition, for understanding of the mixture behaviour in the thermodynamic loop. We currently work on binary mixtures, but ternary mixtures can be studied as well later on. Previous studies have shown that the addition of other chemicals can modify the fluid properties, critical point and equilibrium lines, with sometimes consequences such as reduction of the optimal high pressure or any other way to enhance the coefficient of performance (COP). Mixtures including $CO_2$ have been widely studied in the context of extraction processes, for petroleum, cosmetic or food industrial activities for instance [1], and mostly at higher temperatures. But very few studies are devoted to refrigeration or heat pumps using those mixtures. In the study done by [2,3], $CO_2$ + propane mixtures were used with the conclusion that the refrigerating efficiency may be improved over that of a pure $CO_2$ cycle. Simulations of cycles using dimethyl ether (DME) in addition to $CO_2$ also tend to show performances improvements [4]. $CO_2$ mixtures with HFCs have also been studied in the field of automotive refrigeration [5]. In the latter study, authors showed significant improvement of the refrigeration cycle COP by, for instance, lowering the required pressure ratio of the compressor. This experimental improvement of the COP was not properly understood as simulations based on REFPROP were rather in contradiction to experimental results, and no measurements of the circulating composition were available. In the present study, our aim is to test $CO_2$ mixtures in the field of heat pumps applications, especially hot water production for central heating of buildings, with real-time measurement of the composition of the circulating fluid. First, a quick analysis of the literature will show the expected influence of the addition of other compounds to $CO_2$. Then the experimental bench will be presented, along with the first results for a $CO_2$ + propane and $CO_2$ + R-1234yf mixtures. The detailed methodology for online fluid mixture composition monitoring will take place in the last part of the article.

## 2. $CO_2$ BASED MIXTURES

Adding components to R-744 ($CO_2$) refrigerant will modify its properties and this can be beneficial mostly in two ways. The critical temperature may be moved up so that condensation may occur at temperatures higher than 304 K (31 °C, critical temperature of $CO_2$) leading to lower temperature variations in the high temperature exchanger. Working pressures are also expected to be lower than with pure $CO_2$. This may be illustrated for instance in Figure 1, which shows the critical line and some bubble and dew curves for the $CO_2$ + propane mixture [6]. In Figure 1 we can see that liquid vapor equilibrium takes place for example at 328 K (55 °C) and pressures below 6 MPa. This allows to use a more efficient subcritical cycle for the central heating applications, as they require a lower temperature increase than for domestic hot water production. Of course, it is still possible to operate at higher pressures in a transcritical mode, if greater temperature shifts are required when dealing with domestic hot water production.

In order to increase the critical point, the added compound should be less volatile than $CO_2$ as for example propane (Tc = 369.89 K). But looking at Figure 2 it can be seen that great amounts of propane must be added in order to reach a sufficient critical temperature for our purposes. For example, if we need a mixture critical temperature of say 328 K, then the composition should be of the order of 50 % $CO_2$ only. If we want to reduce the amount of the added compound, an even less volatile compound should be used as for instance ethanol (Tc = 514.71 K). As shown in Figure 2, a composition of nearly 90 % $CO_2$ and only 10 % ethanol will result in the same critical temperature (328 K). However, in the case of ethanol addition, isothermal lines of $CO_2$ are so heavily modified that temperature glides can reach several tens of kelvins during evaporation and condensation [7], and this is an issue. For this particular reason, the architecture of the heat pump would have to be modified so that the added component will circulate through the loop without being stored in the evaporator or in the suction line bottle. When such mixtures are to be tested, previous works can be analyzed, as for example the so-called auto-cascade refrigeration systems [8] or the work of [9].

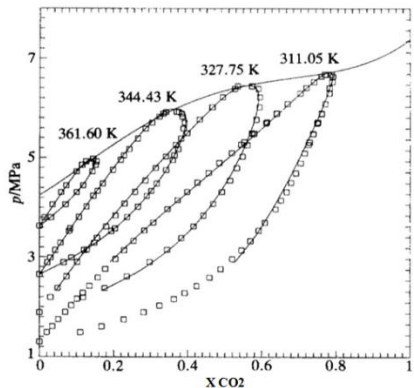

**FIGURE 1.** Bubble and dew curves for $CO_2$ + propane mixtures (Vicky G. Niesen and James C. Rainwater, 1990)

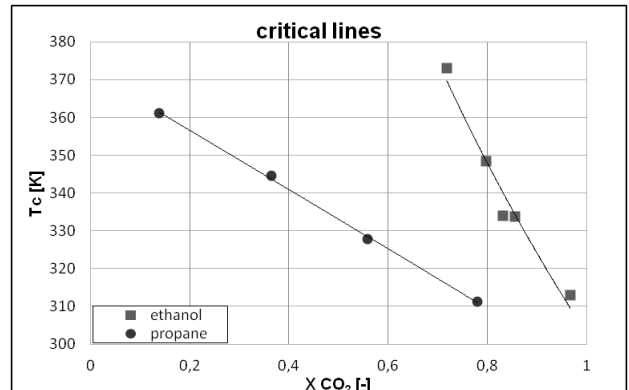

**FIGURE 2.** Critical lines for $CO_2$ + ethanol mixtures (Data from Galicia-Luna and Ortega-Rodriguez, 2000) and $CO_2$ + propane mixtures (Data from Niesen and Rainwater 1990)

First candidates for $CO_2$ based mixtures are presented in Table 1, along with expected effects. The aim is to enhance performances on the high pressure side of the cycle for heat pump applications. All these compounds have critical temperatures above the critical temperature of the $CO_2$ (304.45 K).

TABLE 1. Candidates for mixtures with $CO_2$

| Chemicals families | Species | Critical temperatures [Kelvin] | [°C] | Critical pressures [bar] | NBP [°C] | GWP | Drawbacks |
|---|---|---|---|---|---|---|---|
| Hydrocarbons | Butane | 425.13 | 151.98 | 37.96 | -0.49 | 3 | Flammable |
| | Propane | 369.9 | 96.75 | 42.5 | -42.11 | 20 | Flammable |
| | Ethane | 305.32 | 32.17 | 48.72 | -88.58 | | Flammable |
| | Isobutene | 407.81 | 134.66 | 36.29 | -11.75 | 4 | Flammable |
| | Propene | 364.21 | 91.06 | 45.55 | -47.62 | | Flammable |
| Alcohols | Methanol | 512.6 | 239.45 | 81.04 | 64.48 | Low | Toxic, flammable |
| | Ethanol | 514 | 240,85 | 61.4 | 78.24 | Low | Flammable |
| Ethers | Dimethyl ether | 400.38 | 127.23 | 53.37 | -24.78 | <1 | Toxic, flammable |
| Esters | Methyl acetate | 506.6 | 233.45 | 47.5 | 57-58 | | Irritant, Flammable |
| | Acetone | 508.1 | 234.95 | 47 | 56.01 | | |
| Non-organics | Ammonia | 405.4 | 132.25 | 113.33 | -33.33 | | Toxic, corrosive |
| | $CO_2$ | 304.13 | 30.98 | 73.77 | -78.46 | 1 | |
| HFO | R-1234yf | 367.85 | 94.7 | 33.82 | -29.45 | | Flammable |

# 3. EXPERIMENTAL HEAT PUMP APPARATUS

## 3.1 Bench Description

The experimental facility was designed with modularity in mind. Depending on the investigated mixture, the loop can be modified to create new circuits if necessary. The bench measurements and controls are performed using LabView. In its present state, it is limited to a classical one stage loop. It is based on a water to water $CO_2$ heat pump, built up with commercial R-744 components (Figure 3): variable speed scroll compressor, brazed plate gas cooler, brazed plate evaporator, internal heat exchanger with liquid receiver and an electronic expansion valve. The heat pump loop supplies from 2 to 5 kW of heat output. It is instrumented with thermocouples (carefully recalibrated in our lab), pressure transmitters, two Coriolis flow meters (for mass flow rate and density measurements for both the high and low pressure sides of the cycle), a wattmeter, five in-line flow-cells for near infra-red spectrum measurements and 3 micro-samplers for gas chromatography analysis.

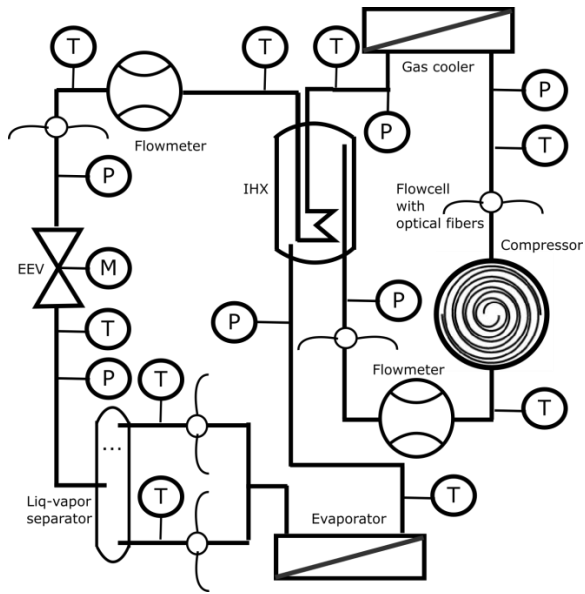 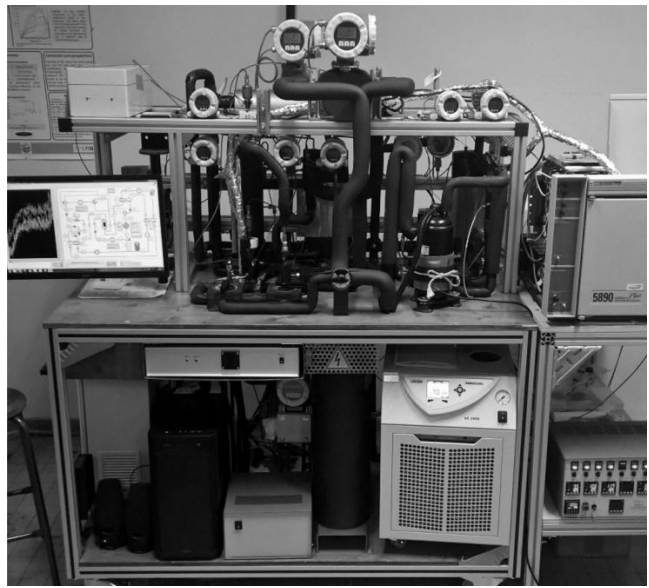

**FIGURE 3.** Heat pump loop and experimental bench

## 3.2 Pure $CO_2$ Heat Pump Characterization

The experimental heat pump is first characterized using $CO_2$ only. Maximum efficiencies are determined for two hot water modes:
- Domestic hot water (DHW) production, when water is heated from 283 to 338 K (10 to 65 °C). COP accuracy is here 1.4 %.
- Central heating (CH) mode, when water is heated from 303 to 308 K (30 to 35 °C). COP accuracy is here 3 %.

For both modes, the brine temperature at the inlet of the evaporator is set at 280 K (7° C).
The cycle works as a conventional $CO_2$ water to water transcritical heat pump. This step allows refining the refrigerant load and control parameters, such as control of the high pressure level through variable speed of the compressor and opening of the electronic expansion valve (EEV) [10]. Optimal high pressure is researched. The results obtained with optimized pure $CO_2$ cycle will be used as a reference for comparison of mixture cycles. Figure 4 shows the performance of the $CO_2$ heat pump loop. Preliminary runs have shown that the amount of $CO_2$ loaded into the loop has almost no influence on the maximum achievable COP, but does have an influence on the operating range of the heat pump. Maximum COP in CH and DHW operating modes are respectively 3.2 and 3.4. These are reached with optimal high pressures of 76 and 92 bar.

## 3.3 Mixtures Evaluation

After characterization of the heat pump loop with pure $CO_2$ refrigerant, binary mixtures are assessed. To load the components of the fluid mixtures into the heat pump loop, it is first vacuum pumped, the less volatile component is loaded, and then $CO_2$ is injected in the loop to complete the refrigerant charge. Performances of the heat pump with mixtures are assessed using the same central heating mode, and domestic hot water mode, as with pure $CO_2$. Optimal high pressure level will be sought to obtain maximum COP. If some COP improvements are noticed, compared to the pure $CO_2$ cycle, the maximum COP is to be researched through variations of the fluid composition and charge.

First results for a $CO_2$ and propane mixture are available in Figure 4.

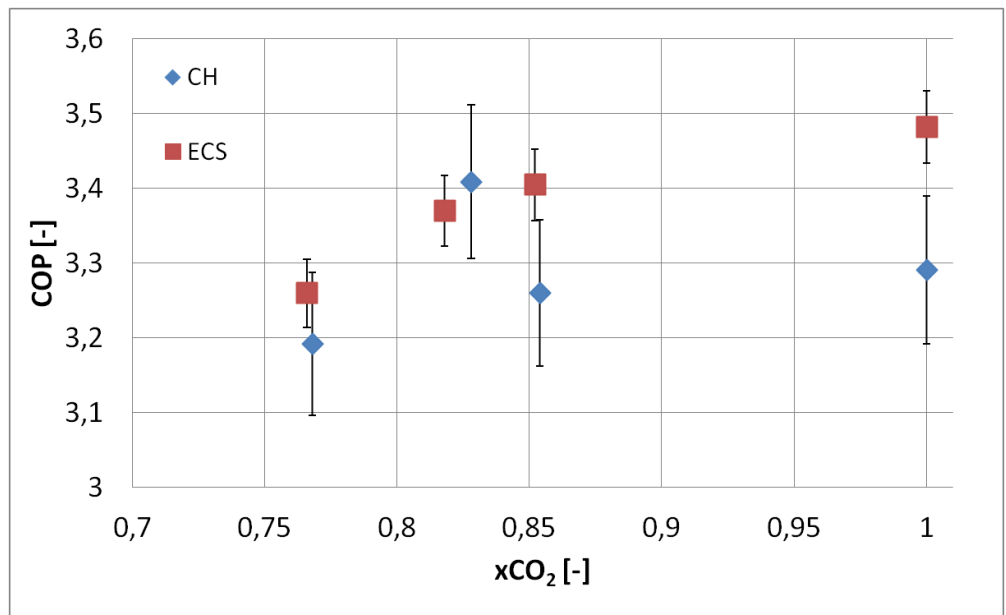

**FIGURE 4.** Maximum COPs of $CO_2$ and propane mixtures

The maximum achievable performance in DHW mode is so far reduced with the addition on propane. Regarding the central heating mode, the performance enhancement recorded with about 17 % of circulating propane remains in limit of the COP measurement accuracy.

Figure 5 shows the performance results for one $CO_2$ & R-1234yf mixture. Circulating molar fractions measured are about 5.5 % of R-1234yf of the DHW mode and 5.0 % for CH mode.
As with addition of propane, addition of R-1234yf reduces the working pressures. For DHW mode, no performance enhancement can be observed, but the CH heating achieved higher COPs with the mixture than pure $CO_2$.

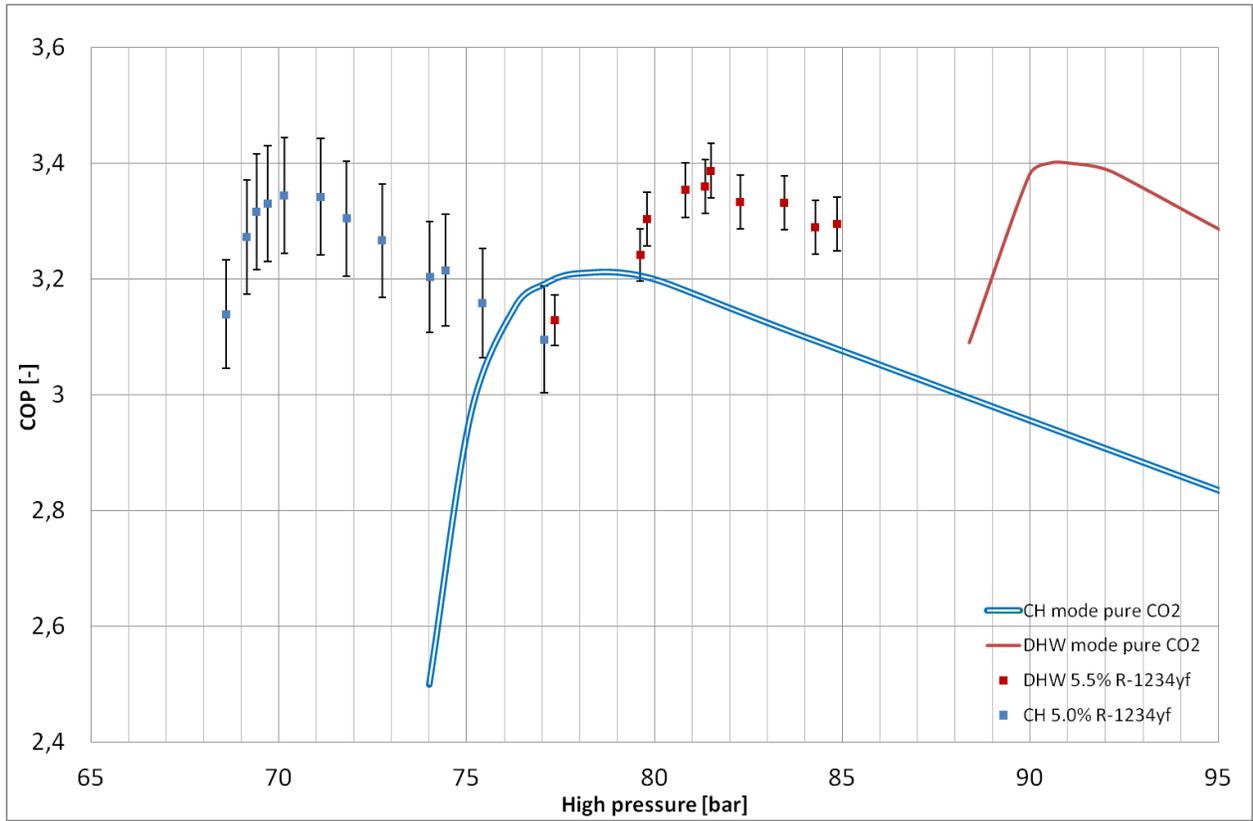

**FIGURE 5.** Performances with a $CO_2$ + R-1234yf mixture compared to pure $CO_2$ performances

# 4. COMPOSITION MEASUREMENTS

The heat pump circuit is filled up with $CO_2$ based mixtures. Overall composition is well known, as the mass of each introduced compound is measured during the process of filling. But to understand the mixture behaviour in the cycle, measurements of the circulating fluid composition are needed. We rely on a non-intrusive in-situ technique which is commonly used in the field of analytical chemistry, namely infrared spectroscopy, coupled with statistical methods known as multivariate calibration [11].

## 4.1 In-line Cells and Optical Setup

The specific difficulty we are facing is the fact that our samples are under varying pressures and temperatures. To address this problem, we use inline flow cells which are tubes equipped with two high pressure windows through which the light passes. Optical fibers are used to lead the light to these optical cells. Figure 5 illustrates the optical setup on the bench. Modulated light coming from the spectrometer is directed to the cell through optical fibers, and an optical multiplexer allows to select the measurement point. The transmitted light is returned to the spectrometer through optical fibers and a multiplexer. Specific infrared wavelengths are absorbed, depending on the chemicals and their concentrations. It results in absorption spectra, which will be analyzed by a chemometric method.

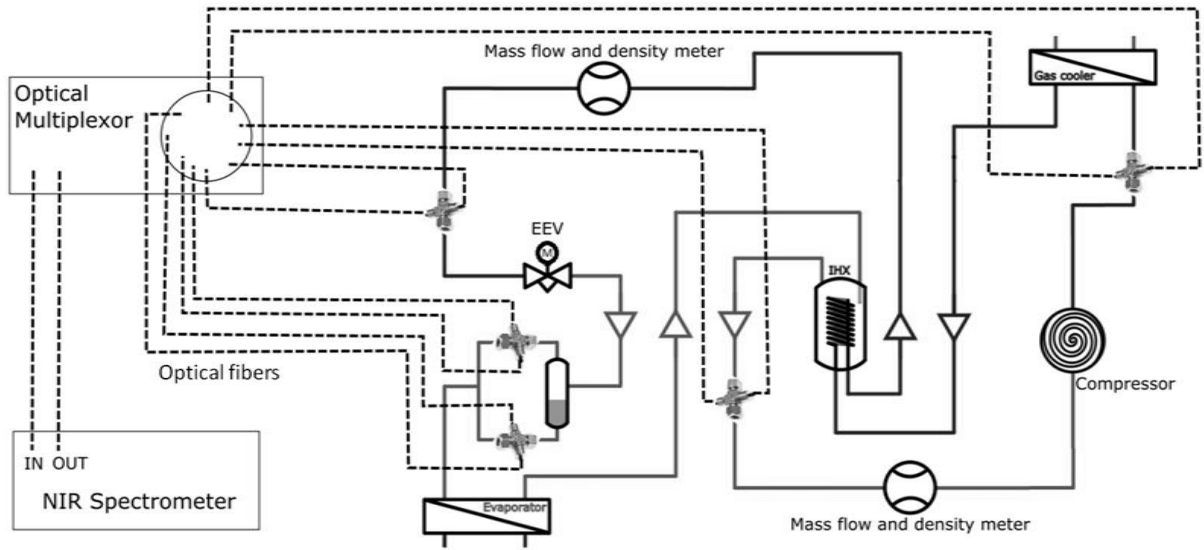

**Figure 6.** Optical setup for NIR absorption spectrum acquisitions

Five flowcells are settled on the loop (see Figure 6), where the refrigerant is supposed to be a homogeneous phase:
- Discharge line at the compressor outlet.
- Expansion valve input.
- Two flowcells at the expansion valve output (one for the liquid phase, the other for the gas phase).
- Compressor inlet.

Because of different volatilities of the chemicals, we expect that during unsteady phases, the less volatile component will accumulate in the compressor's sump, in the exchangers, or in the suction line bottle. Also, refrigerant absorption in oil modifies refrigerant mixtures circulating compositions. That is why circulating composition is not equal to the introduced mixture composition. The composition measurement method using five flowcells will help to determine the distribution of the different species among the parts of the loop.

## 4.2 Chemometric Methods

Chemometric methods allow real-time composition measurement by analysis of the absorption spectrum (Figure 7). They consist in processing the recorded spectrum of a fluid with unknown composition, using statistical methods like partial least squares (PLS) or principal components regression (PCR) [11]. However, it is not a direct measurement method. It should first be carefully calibrated by recording a number of spectra for mixtures of known composition varying in the range of expected compositions and concentrations. These are known as the calibration set or the training set.

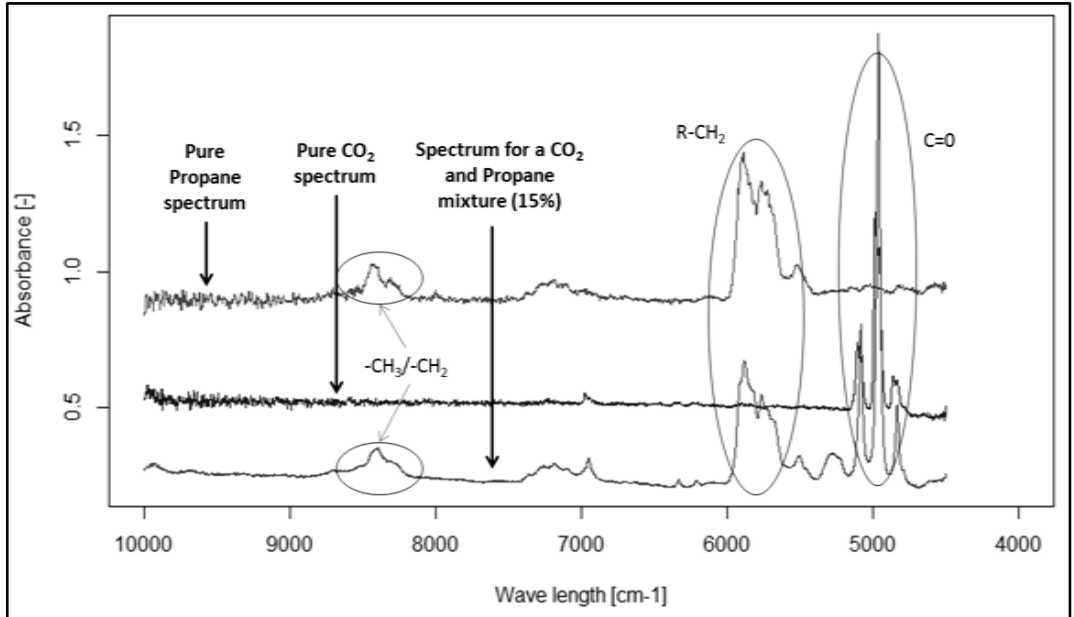

**FIGURE 7.** Absorption spectra for $CO_2$ + propane and bands characterization according to [12]

A specific difficulty in our application of chemometrics is that, for given pressure and temperature, not all compositions exist. It is possible to reproduce (P, T) conditions in equilibrium cells but this would have been too time-consuming. Instead, in order to generate our calibration set, we added micro-samplers next to the flow cells to perform a composition measurement of the fluid mixture by gas chromatography. Micro-samplers are small heated electronic valves that open during a defined time length. A small amount of mixture leaks (and vaporizes if the sample is liquid) and flows into a heated transfer line in which the vector gas (here helium) flows. The sample is then carried to the gas chromatograph column for components separation, and then chemicals are detected by a thermal conductivity detector (TCD). The signal generated by this detector draws peaks for each compound and the peak area is related to the number of moles of the component in the sample. The gas chromatograph is itself calibrated to determine the relationship between peaks areas and mole numbers, for each chemical species.

Prior to any COP measurement with a given mixture, a series of runs is performed only to generate a calibration or training set of spectra with known molar fractions and concentrations. Figure 8 shows pressure, temperature and concentration range used so far. Once the heat pump loop is stabilized, spectra are recorded and samples of the mixtures are taken and analysed by chromatography.
Note that the spectroscopic method deals with concentrations of the species while the chromatographic technique only gives access to molar fractions. Hence, the density of the fluid is necessary to relate these quantities.

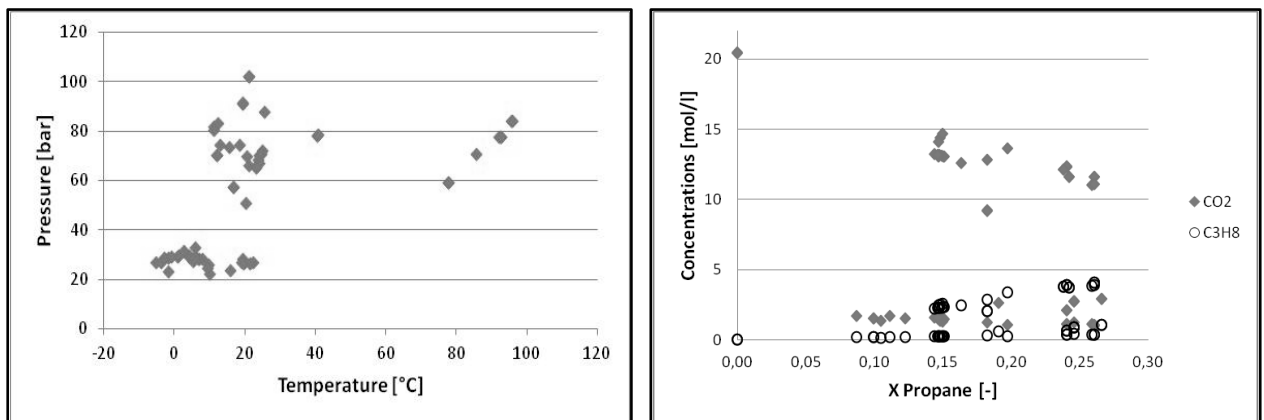

**FIGURE 8.** 49 Samples considered for training set, with propane molar fractions from 0 to 0.266

To quickly assess the chemometric method, a first training set of spectra (Figure 8), with known concentrations, has been processed using the quantification software supplied with the NIR spectrometer. For now, the quantification software can be used to generate models for $CO_2$ and propane concentrations. To do so, the training set is fed to a Partial Least Square (PLS) algorithm, with a cross validation method. Figure 9 shows that calculated concentrations fits very well the measured ones.

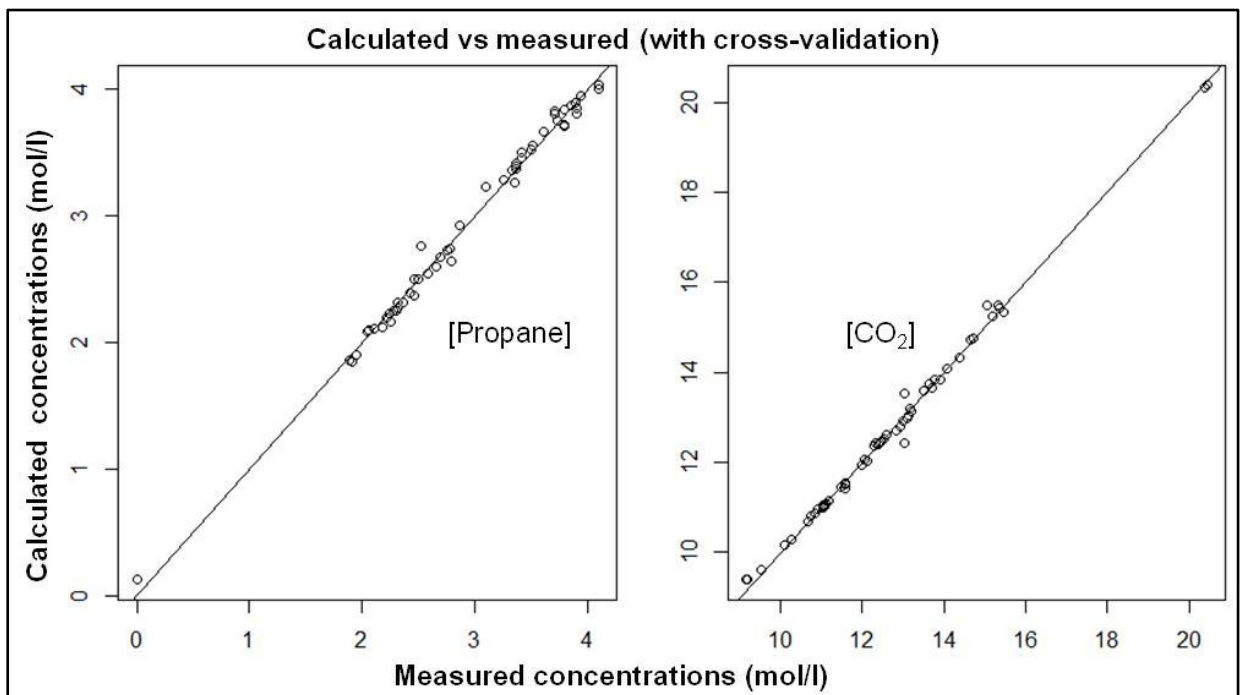

**FIGURE 9.** Validation of the concentration models.

Maximum error of concentration predictions are 1.5 % for $CO_2$ and 3.5 % for propane in the explored concentration ranges.

## 6. CONCLUSIONS

Characterization of the heat pump has been carried out with pure $CO_2$. $CO_2$ and propane mixtures have shown interesting potential for performance enhancements in central heating mode. The calibration of the chemometric method for these binary mixtures has been successfully performed and can still be improved. It already allows us to relate the evolution of the performances to the measured circulating compositions.

This is a first step in a project aiming at the study of $CO_2$ based mixtures. The addition of little amounts of HFCs, hydrocarbons and other derivatives will be carried out and assessed. Obtained efficiencies will be compared to the pure $CO_2$ refrigerant cycle performance measured on this apparatus. $CO_2$ and R-1234yf mixtures are currently under testing, and already show very promising performances.